\documentstyle[sprocl]{article}
\input{psfig}

\def\Journal#1#2#3#4{{#1} {\bf #2}, #3 (#4)}

\def\NPB{{\em Nucl. Phys.} B}
\def\PLB{{\em Phys. Lett.}  B}
\def\PRL{\em Phys. Rev. Lett.}
\def\PRD{{\em Phys. Rev.} D}

\def\be{\begin{equation}}
\def\ee{\end{equation}}
\def\bea{\begin{eqnarray}}
\def\eea{\end{eqnarray}}

\newcommand{\Eq}[1]{Eq.({\ref{#1}})}
\newcommand{\Fig}[1]{Fig.\protect\ref{#1}}
\newcommand{\alt}{\raisebox{.6ex}{$<$} \hspace{-.8em} \raisebox{-.6ex}{$\sim$}}
\newcommand{\Ndf}{N_{\rm{DF}}}
\newcommand{\Nsmear}{N_{\rm{smear}}}

\newcommand{\Tate}{\rule{0cm}{1.1em}}
\newlength{\Tatescale}
\newcommand{\Hs}{\hspace*{1em}}
\newcommand{\Bs}{\hspace*{-0.5em}}

\begin{document}
\title{Glueball properties in anisotropic SU(3) lattice QCD
with improved action}
\author{Noriyoshi ISHII}
\address{
The Institute of Physical and Chemical Research(RIKEN),\\
2-1 Hirosawa, Wako, Saitama 351-0198, JAPAN}

\author{Hideo SUGANUMA}
\address{Tokyo Institute of Technology,\\
2-12-1 Ohkayama, Meguro, Tokyo 152-8552, JAPAN}

\author{Hideo MATSUFURU}
\address{Yukawa Institute for Theoretical Physics, Kyoto University,\\
Kitashirakawa-Oiwake, Sakyo, Kyoto 606-8502, JAPAN}

\maketitle\abstracts{  We  study  the  glueballs  properties  at  finite
temperature using  $SU(3)$ lattice  QCD at the  quenched level  with the
anisotropic lattice.   We use the tree-level  Symanzik $O(a^2)$ improved
action.  We  present our  preliminary  results  which  shows the  slight
reduction of the scalar glueball mass near $T_c$.  }

\section{Introduction}
The $SU(3)$ lattice QCD Monte Carlo simulations suggest the existence of
the  finite   temperature  deconfinement  transition   at  the  critical
temperature $T_c  \simeq 260$ MeV at the  quenched level \cite{karsch1}.
At the  quenched level,  without involving dynamical  quark excitations,
the lightest  physical excitation is  the scalar glueball with  the mass
$1.5$    GeV     $\alt\,    m_{    \rm    G}\,     \alt\,    1.7$    GeV
\cite{morningstar,weingarten},  which can  be expected  to  dominate the
thermodynamic  properties  below $T_c$.   
The  closed  packing model,  a
phenomenological model for the QCD phase transition, provides us with an
intuitive picture that the  deconfinement transition takes place when an
``hadrons'' excited from the  thermal system begin overlap significantly
with one  another.  However, since the contributions  from the glueballs
are  suppressed by  a rather  strong statistical  factor  as $e^{-m_{\rm
G}/T}$, we may wonder if the total number of the glueballs to be excited
may be insufficient  in the quenched QCD.  
To keep  the consistency with  the closed packing model,  the resolution
may be found in either a  tremendous expansion of the glueball size or a
drastic reduction of the mass
\footnote{The reduction  of a glueball  mass is also anticipated  in the
monopole  condensation picture of  the confinement  \cite{ichie}.}  near
the critical temperature (See \Fig{cpm}).  Indeed, the lattice result of
the correlator of Polyakov loops supports the change in the shape of the
$q\bar{q}$ potential at finite temperature \cite{matsufuru,karsch2}.  As
a consequence,  the phenomenological  potential model suggests  also the
change in  the wave function of  the glueball particularly  its mass and
size.
It is thus  important to  study the  properties of  glueballs at
finite temperature,  although no such studies  have been done  so far on
the lattice.   Even for  typical hadrons, the  most studies of  the mass
shift on the lattice at  finite temperature are unfortunately limited to
the  screening mass  \cite{}, i.e.,  the correlation  along  the spatial
direction \cite{detar-kogut}, whereas what  is really interesting is the
direct informations of the mass from the correlations along the temporal
direction.  This is because the lattice points available in the temporal
direction decrease as the temperature becomes higher, and the extraction
of the mass  spectrum becomes more difficult.  
Recently, the use of the anisotropic lattice, where the temporal spacing
is    finer   than    the    spatial   one,    has   been    established
\cite{matsufuru,klassen}, which enables  us to overcome the difficulties
mentioned above.  In this paper, after analytical consideration with the
closed  packing  model,  we  will  outline how  to  study  the  glueball
properties at finite temperature by means of the anisotropic lattice and
the  smearing method.  We  will present  preliminary results  in $SU(3)$
lattice QCD with the tree-level Symanzik $O(a^2)$ improved action.
\section{The closed packing model for glueballs in the quenched QCD}
\begin{figure}
\begin{center}
\parbox{0.5\textwidth}{\psfig{figure=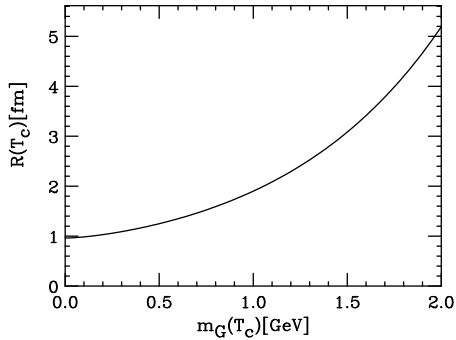,width=0.5\textwidth,angle=90}}
\end{center}
\caption{The  glueball radius $R(T_c)$  so as  to reproduce  $T_c \simeq
260$ MeV  in the closed packing  model is plotted  against the effective
thermal glueball mass $m_{\rm G}(T_c)$.  }
\label{cpm}
\end{figure}
In  this section,  we analytically  investigate the  deconfinement phase
transition  at the  quenched  level  in terms  of  the low-lying  scalar
glueball  properties.  Consider  the  bag model  picture  of the  scalar
glueball.  In  this picture,  a sphere of  radius $R$  distinguishes the
outside from the  inside, where the gluonic degrees  of freedom appears.
As long  as the temperature $T$  is low, the number  $N(T)$ of thermally
excited  bags is  small.   However, as  the  temperature increases,  the
number $N(T)$ of  bags increases.  Finally, these bags  begin to overlap
each other and  cover the whole space, and  thus the deconfinement phase
transition takes  place in  the closed packing  model. The ratio  of the
volume inside the gluonic bags to the space volume $V$ is given as
\begin{equation}
	r_V(T)
=
	{{4\pi\over 3} R(T)^3 \times N(T) \over V}
=
	{4\pi \over 3} R(T)^3
	\int {d^3k\over(2\pi)^3}
	{1\over e^{E(\vec k,T)/T} - 1}.
\end{equation}
Now  we simplify  the  dispersion relation  to  the quasi-free  relation
$E(\vec k,T)  = \sqrt{\vec  k^2 + m^2_{\rm  G}(T)}$ with  $m_{\rm G}(T)$
begin  the  effective  glueball  mass  at temperature  $T$.   Since  the
deconfinement phase  transition is expected  to take place  when $r_V(T)
\simeq  1$, the glueball  radius $R(T_c)$  near the  deconfinement phase
transition is estimated as
\begin{equation}
	R(T_c)
\simeq
	\left(
		{4\pi \over 3}
		\int {d^3k\over(2\pi)^3}
		{1\over e^{\sqrt{\vec k^2 + m^2_{\rm G}(T_c)}/T_c} - 1}
	\right)^{-1/3}.
\end{equation}
In  \Fig{cpm}, the  glueball  radius  $R(T_c)$ so  as  to reproduce  the
deconfinement  phase  transition at  $T_c=260$  MeV  is  plotted as  the
function of  the glueball  mass $m_{\rm G}(T_c)$  in the  closed packing
model.  Thus,  if $m_{\rm  G}$ persists  to be 1.7  GeV near  $T_c$, the
glueball should have the abnormally  huge radius 3.8 fm. Instead, if $R$
remains to be a typical hadron size about 1 fm, the glueball mass should
reduce drastically as $m_{\rm G}(T_c) \alt 0.5$ GeV.
%
%
\section{The smearing method in {\boldmath $SU(3)$} lattice QCD}
Using $SU(3)$ lattice QCD, we study the glueball correlator
\begin{equation}
	G(t)
\equiv
	\langle \tilde O(t) \tilde O(0) \rangle,
\Hs
	\tilde O(t)  \equiv O(t) -  \langle O  \rangle,
\Hs
	O(t)  \equiv \sum_{\vec x}  O(t,\vec x).
\end{equation}
The subtraction  of the  vacuum expectation value  is necessary  for the
scalar glueball with $J^{\rm PC}  = 0^{++}$. The summation over $\vec x$
physically means the  momentum projection as $\vec p =  \vec 0$.  In the
case     of     the      scalar     glueball,     the     summand     is
\cite{morningstar,weingarten,rothe,montvey}
\begin{equation}
	O(t,\vec x)
\equiv
	O_{12}(t,\vec x) + O_{23}(t,\vec x)  + O_{31}(t,\vec x),
\end{equation}
where $O_{ij}(t,\vec x)$ is the real  part of the trace of the plaquette
operator  which  lies  within  the  $i$-$j$  plain.  With  the  spectral
representation, $G(t)$ is expressed as as
\begin{equation}
	G(t)/G(0)
=
	\sum_{n=1}^{\infty} c_n e^{-E_n t},
\Hs
	c_n
\equiv
	\left|\Tate \langle n | \tilde O | 0 \rangle \right|^2
	/ G(0),
\Hs
	G(0)
=
	\sum_{n=1}^{\infty}
	\left|\Tate \langle n | \tilde O | 0 \rangle \right|^2,
\end{equation}
where $E_n$ denotes the energy  of the $n$-th excited state $|n\rangle$.
(In this paper, $|0\rangle$  denotes the vacuum, and $|1\rangle$ denotes
the  ground-state $0^{++}$  glueball.)  Note  that $c_n$  is  a positive
number with $\sum  c_n = 1$, and thus $G(t)/G(0)$ can  be expressed as a
weighted average of the exponentials $e^{-E_n t}$ with the weight $c_n$.
On a finite lattice with  the spacing $a$, the simple plaquette operator
$O_{ij}(t,\vec  x)$ has  a small  overlap with  the glueball  ground state
$|G\rangle \equiv |1\rangle$, and  the extracted mass from $G(t)$ looks
heavier owing  to the  excited-state contamination.  This  small overlap
problem  originates  from that  the  plaquette  operator  has a  smaller
``size'' of $O(a)$ than the physical size of the glueball.  This problem
becomes severer near the continuum limit, i.e., $a \simeq 0$.
We thus have  to enhance the overlap by  improving the glueball operator
so as  to have approximately the same  size as the physical  size of the
glueball.  One of the systematic ways to achieve this is known to be the
smearing  method  \cite{morningstar,rothe,ape,takahashi}.  The  smearing
method is expressed as the iterative replacement of the original spatial
link  variables $U_i(s)$  by the  associated fat  link  variables, $\bar
U_i(s) \in SU(3)_c$. The fat link variable $\bar U_{i}(s)$ is defined to
be the $SU(3)_c$ element which maximizes
\begin{equation}
	\mbox{Re}
	\mbox{Tr}
	\left\{
		\bar U_i^{\dagger}(s)
		\left[
			\alpha U_i(s)
		+	\sum_{j\neq i}
			\sum_\pm
			U_j(s)
			U_i(s\pm\hat j)
			U_j^{\dagger}(s\pm\hat i)
		\right]
	\right\},
\label{fat.link}
\end{equation}
where $U_{-\mu}(s) \equiv  U^{\dagger}_{\mu}(s - \hat\mu)$, and $\alpha$
is a real parameter.  A schematic  illustration of the fat link is given
as
\begin{equation}
\parbox{0.15\textwidth}{\psfig{figure=ishii.1.eps,height=0.12\textheight,%
bbllx=110bp,bblly=530bp,bburx=180bp,bbury=740bp}}
\sim
\alpha\times
\parbox{0.15\textwidth}{\psfig{figure=ishii.2.eps,height=0.12\textheight,%
bbllx=110bp,bblly=530bp,bburx=180bp,bbury=740bp}}
+
\parbox{0.4\textwidth}{\psfig{figure=ishii.3.eps,height=0.12\textheight,%
bbllx=40bp,bblly=590bp,bburx=247bp,bbury=808bp}}
+
\parbox{0.4\textwidth}{\psfig{figure=ishii.4.eps,height=0.12\textheight,%
bbllx=40bp,bblly=590bp,bburx=247bp,bbury=808bp}}
\end{equation}
Note that the summation is only  over the spatial direction to avoid the
nonlocality  in the temporal  direction.  Note  also that  $\bar U_i(s)$
holds the  same gauge transformation properties with  $U_i(s)$.  We will
refer to  the fat link defined  in \Eq{fat.link} as the  first fat link.
The  $n$-th fat link  is defined  iteratively, i.e.,  it is  obtained by
repeating the procedure in \Eq{fat.link} with $(n-1)$-th fat links.  For
the physically  extended glueball operator,  we use the  $n$-the smeared
operator, the  plaquette operator which  is constructed with  the $n$-th
fat link variables.
To  attribute  a  size  to  the  $n$-th smeared  operator,  we  use  the
linearization and the continuum  approximation, and obtain the diffusion
equation \cite{preparation} as
\begin{equation}
	{\partial\over \partial n}
	\phi_i(n; \vec x)
=
	D \triangle \phi_i(n;\vec x),
\Hs
	D \equiv {a_s^2\over \alpha  + 4},
\Hs
	\mbox{$a_s$ : spatial lattice spacing}.
\end{equation}
Here $\phi_i(n;\vec  x)$ denotes the distribution function  of the gluon
field  $A_i(\vec x,t)$  after the  $n$ smearing  operations.  The $n$-th
smeared plaquette  located at  the origin $\vec  x = \vec  0$ physically
means the Gaussian extended glueball operator with the distribution as
\begin{equation}
	\phi_i(n;\vec x)
=
	{1\over (4\pi D n)^{3/2}}
	\exp\left[ - {\vec x^2 \over 4 D n} \right].
\end{equation}
Hence the size of the operator is estimated as
\begin{equation}
	\sqrt{\langle \vec x^2 \rangle}
=
	\sqrt{6 D n}
=
	a_s \sqrt{6n \over \alpha + 4}.
\label{size}
\end{equation}
So far, the smearing method is introduced to carry out the accurate mass
measurement by maximizing the overlap.   However, it can be also used to
give a rough  estimate of the physical glueball size.   In fact, once we
obtain the maximum overlap with some $n$ and $\alpha$, the physical size
of the glueball is estimated with \Eq{size}.
\section{The {\boldmath $SU(3)$} lattice QCD result for glueball mass
and size at {\boldmath $T\neq 0$}}
%
In this  section, we  present our preliminary  results. To  generate the
gauge  field configurations,  we  use the  tree-level Symanzik  $O(a^2)$
improved action \cite{symanzik}
\begin{equation}
	S_{\rm G}
=
\Bs
\begin{array}[t]{l}	\displaystyle
	\beta {1\over \gamma_{\rm G}}
	\sum_{s,i<j\le 3}
	\left[\Tate
		c_{11}(1 - P_{ij}(s)) + c_{12}(2 - R_{ij}(s) - R_{ji}(s))
	\right]
\\\displaystyle
+
	\beta \gamma_{\rm G}
	\sum_{s,i\le 3}
	\left[\Tate
		c_{11}(1 - P_{i4}(s)) + c_{12}(2 - R_{i4}(s) - R_{4i}(s))
	\right],
\end{array}
\end{equation}
where $c_{11}=5/3$,  $c_{12}=-1/12$, $\beta=4.56$ and  $\gamma_{\rm G} =
3.45$.   $P_{\mu\nu}(s)$  is  the  plaquette and  the  rectangular  loop
$R_{\mu\nu}(s)$   is   defined  as   $   R_{\mu\nu}(s)   =  {1\over   3}
\mbox{Re}\mbox{Tr} \left(\Tate\right.   $ $ U_{\mu}(s)  U_{\mu}(s + \hat
\mu)  U_{\nu}(s + 2\hat  \mu) U^{\dagger}_{\mu}(s  + \hat\nu  + \hat\mu)
U^{\dagger}_{\mu}(s +  \hat\nu) U^{\dagger}_{\nu}(s) \left.\Tate \right)
$.  The numerical calculations are performed on the lattice of the sizes
$16^2 \times 24 \times N_t$, $N_t =  28$ ($T = 0.87 T_c$) and $N_t = 96$
($T \simeq 0$)  \cite{matsufuru,klassen}.  These parameters generate the
spatial  lattice spacing as  $a_s =  0.12$ fm  and the  temporal lattice
spacing as $a_t = 0.03$ fm.  We have used 500 gauge field configurations
for $N_t=28$ and 150 for $N_t=96$.  Throughout this section, we take the
smearing parameter as $\alpha = 2.3$.

\begin{figure}
\begin{tabular}{cc}
\psfig{figure=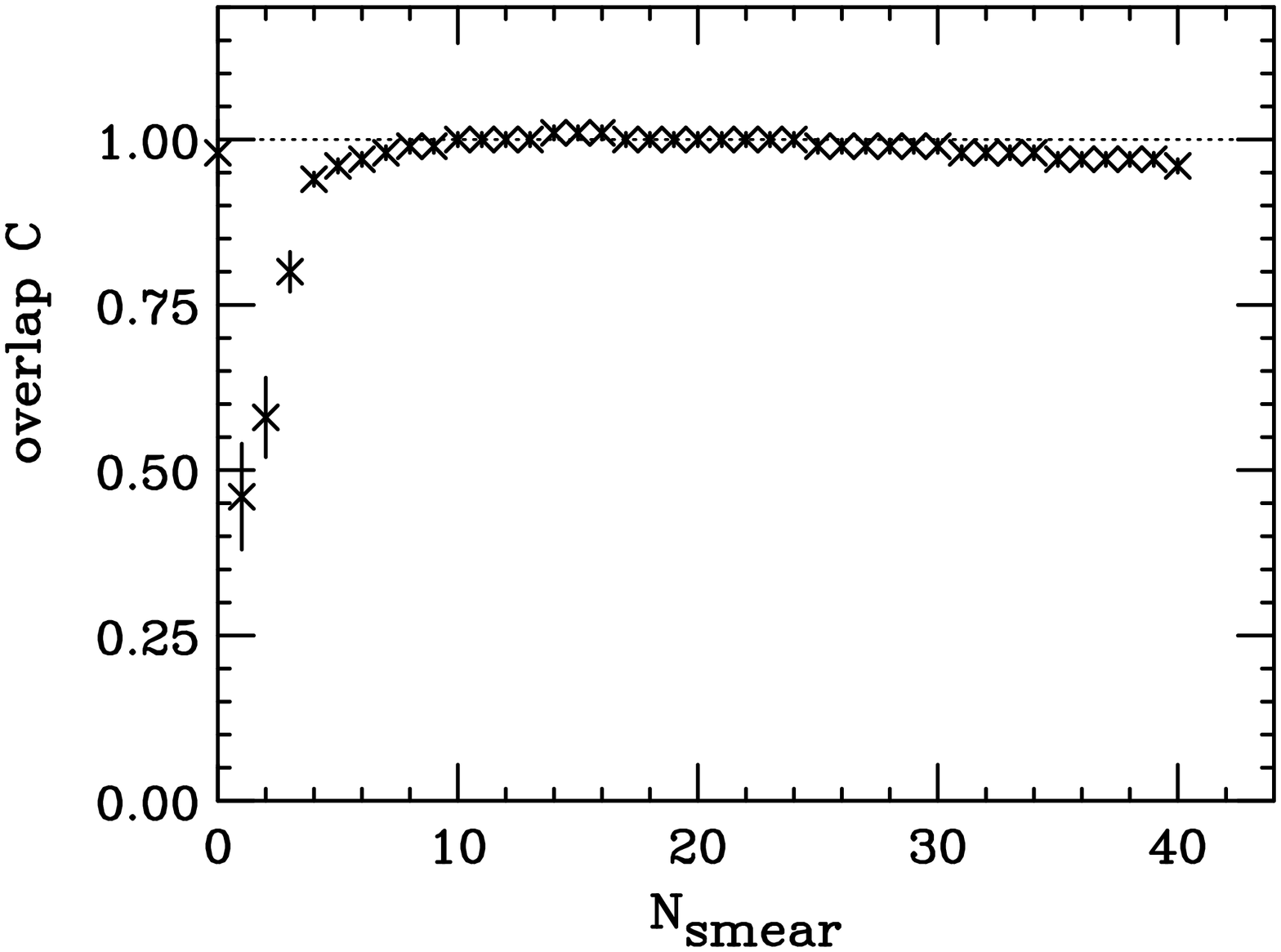,width=0.47\textwidth,angle=90}
&
\psfig{figure=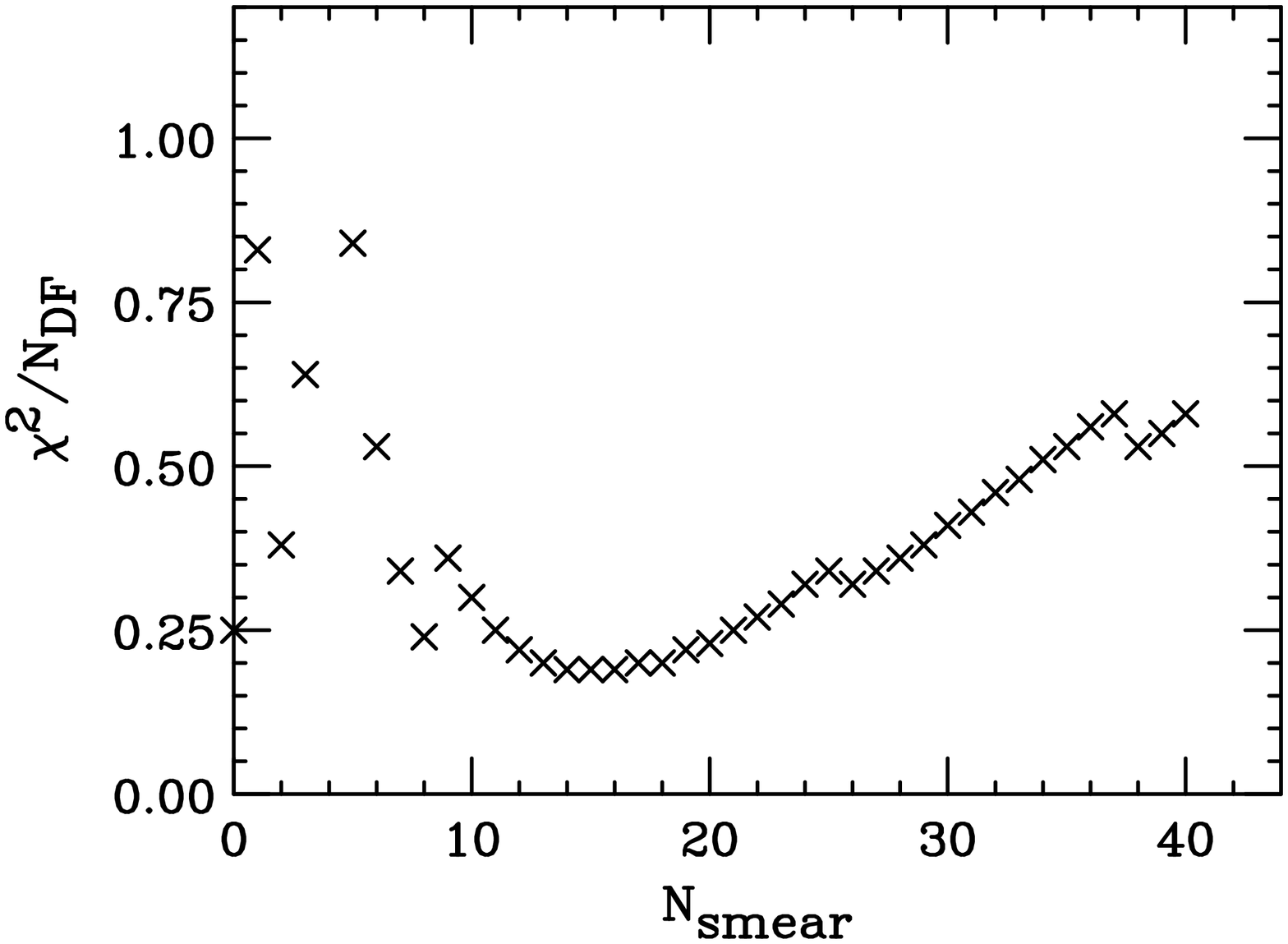,width=0.47\textwidth,angle=90}
\end{tabular}
\caption{The  overlap $C$ is  plotted against  $\Nsmear$ at  $T=0$ (left
figure).   $\chi^2/\Ndf$ is  plotted against  $\Nsmear$ at  $T=0$ (right
figure).}
\label{overlap.chi2.mass.t=0}
\end{figure}

\begin{figure}
\begin{tabular}{cc}
\psfig{figure=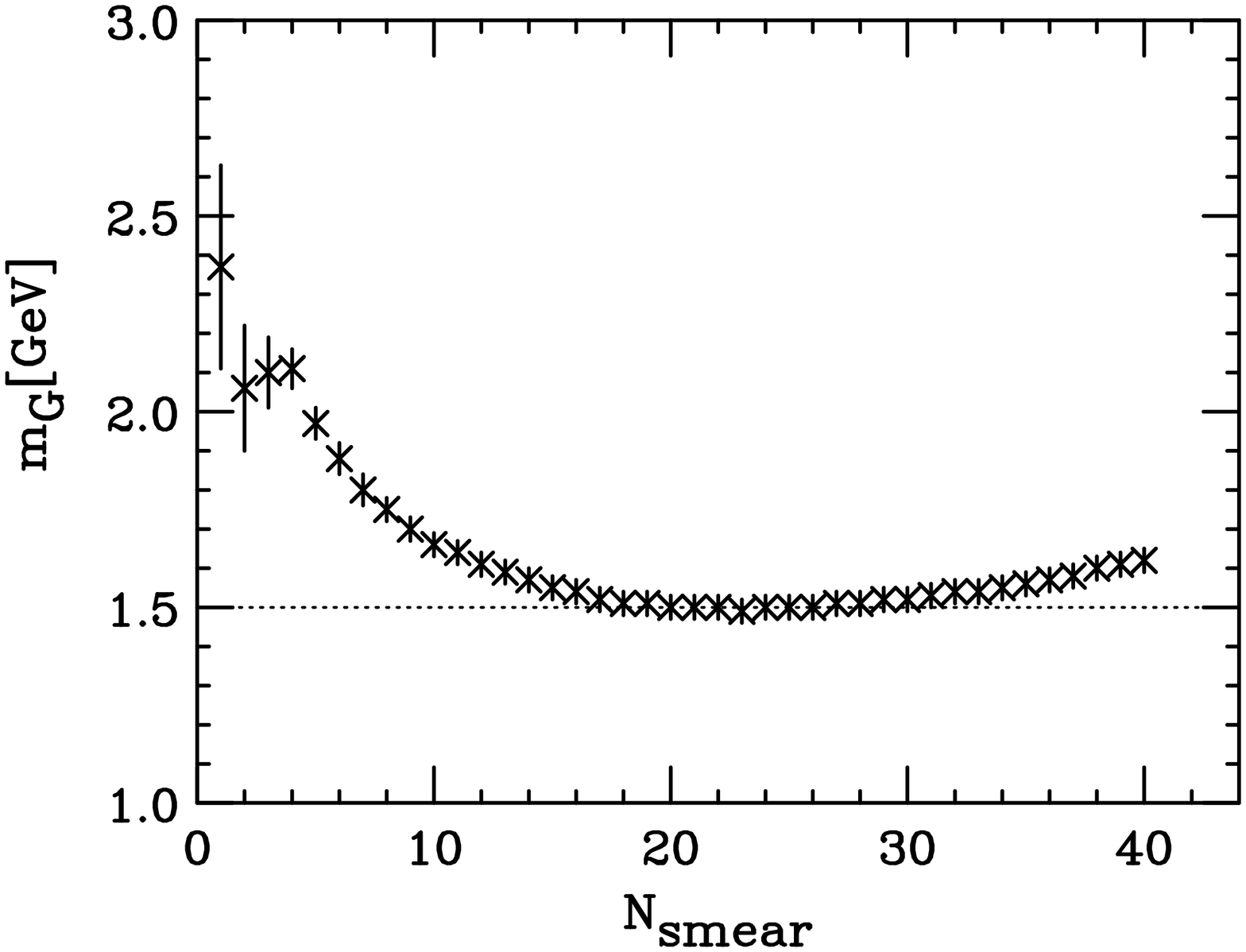,width=0.47\textwidth,angle=90}
&
\psfig{figure=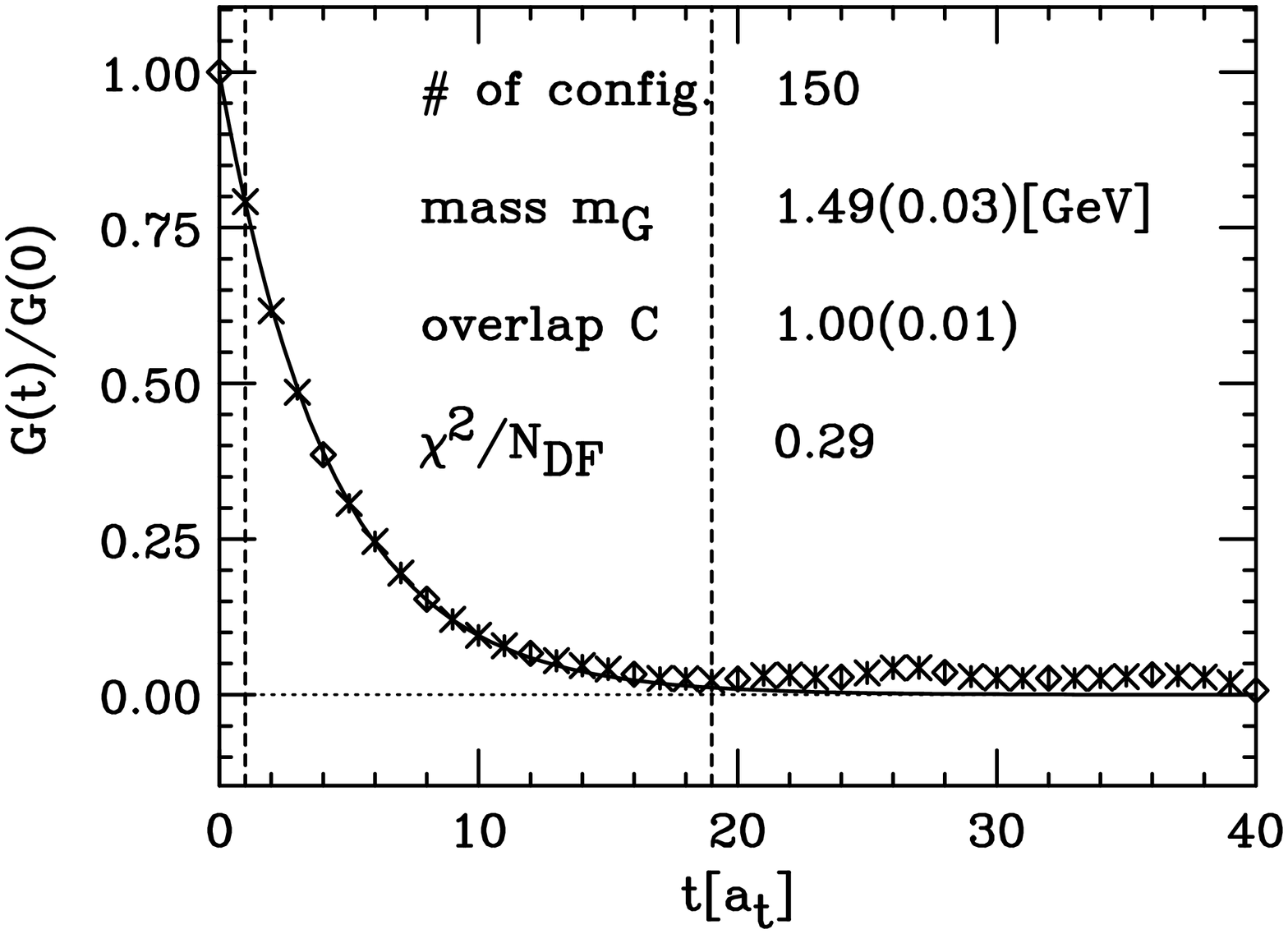,width=0.47\textwidth,angle=90}
\end{tabular}
\caption{The  glueball mass  $m_{\rm  G}$ is  plotted against  $\Nsmear$
(left figure).  The normalized  glueball correlator $G(t)/G(0)$ at $T=0$
is plotted at  $\Nsmear=23$ (right figure).  The curve  denotes the best
single-exponential  fit  from the  region  between  the vertical  dotted
lines.}
\label{after.smearing.t=0}
\end{figure}

In    \Fig{overlap.chi2.mass.t=0},   the    the   overlap    $C   \equiv
\left|\Tate\langle G|\tilde  O|0\rangle\right|^2/G(0)$ and $\chi^2/\Ndf$
are  plotted against  the  iteration number  of  the smearing  $\Nsmear$
($=n$).   The  overlap  becomes  maximum  with  reasonable  $\chi^2$  as
$\chi^2/\Ndf \alt  1$ in the  region $10 \le  \Nsmear \le 24$.   In this
region,  as is  shown  in \Fig{after.smearing.t=0},  the glueball  mass,
which is extracted with the single-exponential fit, takes its minimum as
$m_{\rm G} = 1.49$ GeV  at $\Nsmear=23$.  This indicates the achievement
of the ground-state glueball.  We thus  obtain $m_{\rm G} = 1.49$ GeV at
$T=0$.
The glueball size $R$ is estimated with \Eq{size} as $0.37$ fm $\,\alt\,
R \,\alt\, 0.57$  fm from the maximal overlap region  as $10 \le \Nsmear
\le 24$.   We show also  the glueball correlator  $G(t)$ at $T=0$  for a
smearing number $\Nsmear =  23$ in \Fig{after.smearing.t=0}.  Note that,
if we used the isotropic lattice, we could only use the points indicated
by the diamond symbols, which would be inefficient to get the mass.

\begin{figure}
\begin{tabular}{cc}
\psfig{figure=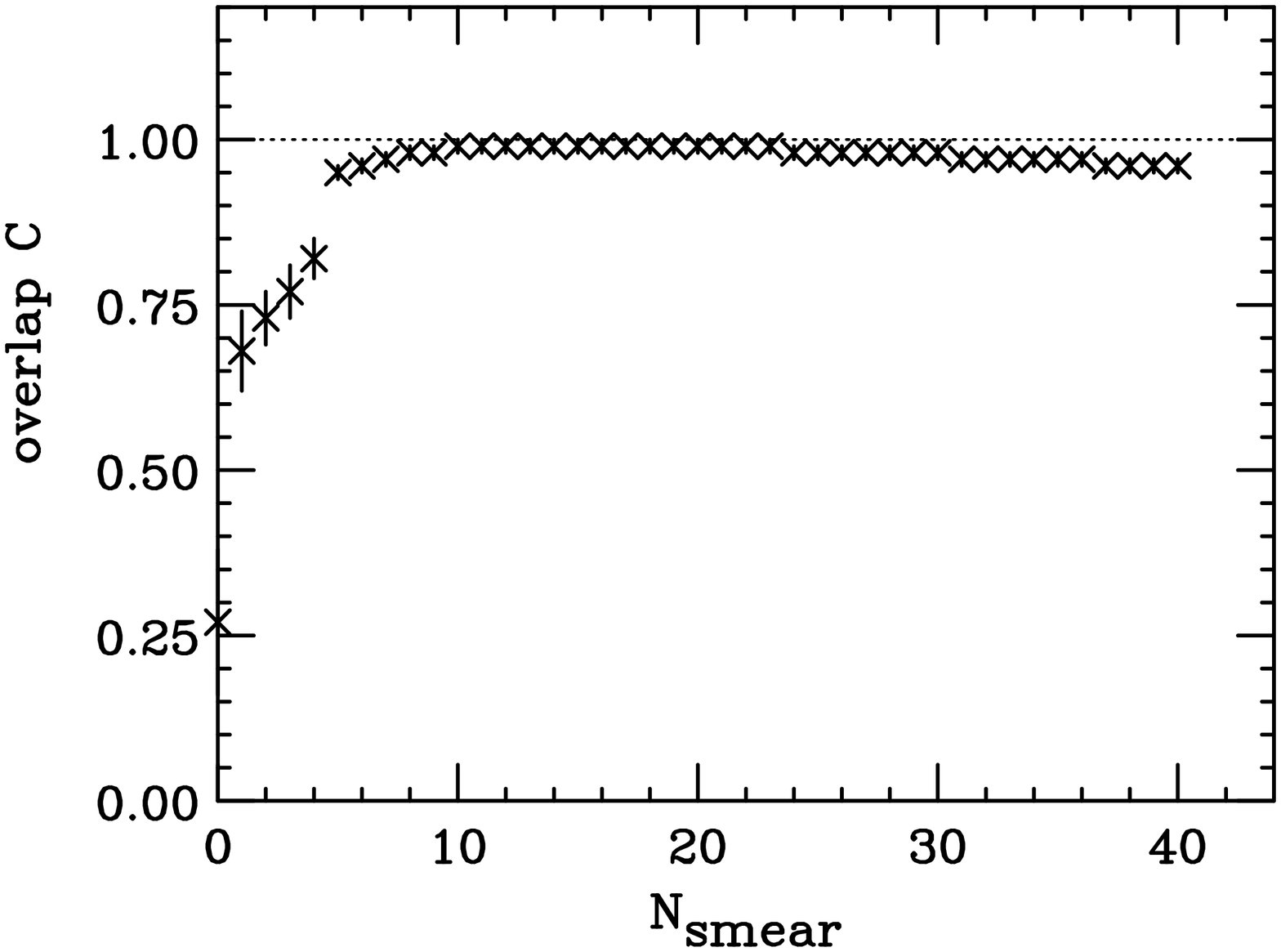,width=0.45\textwidth,angle=90}
&
\psfig{figure=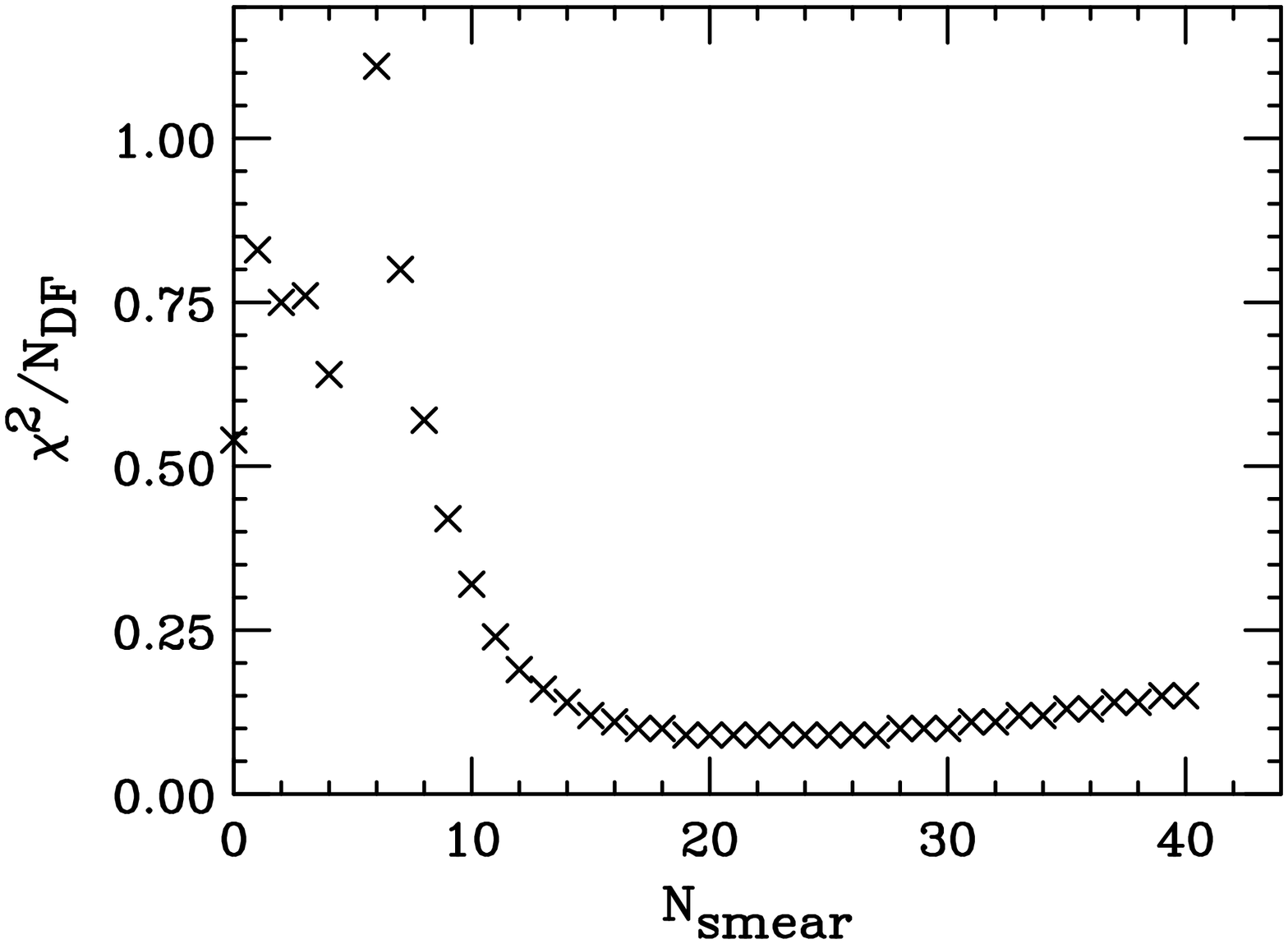,width=0.45\textwidth,angle=90}
\end{tabular}
\caption{The  overlap $C$  is plotted  against $\Nsmear$  at $T=0.87T_c$
(left   figure).   $\chi^2/\Ndf$   is  plotted   against   $\Nsmear$  at
$T=0.87T_c$ (right figure).}
\label{overlap.chi2.region.t=0.87tc}
\end{figure}

\begin{figure}
\begin{tabular}{cc}
\psfig{figure=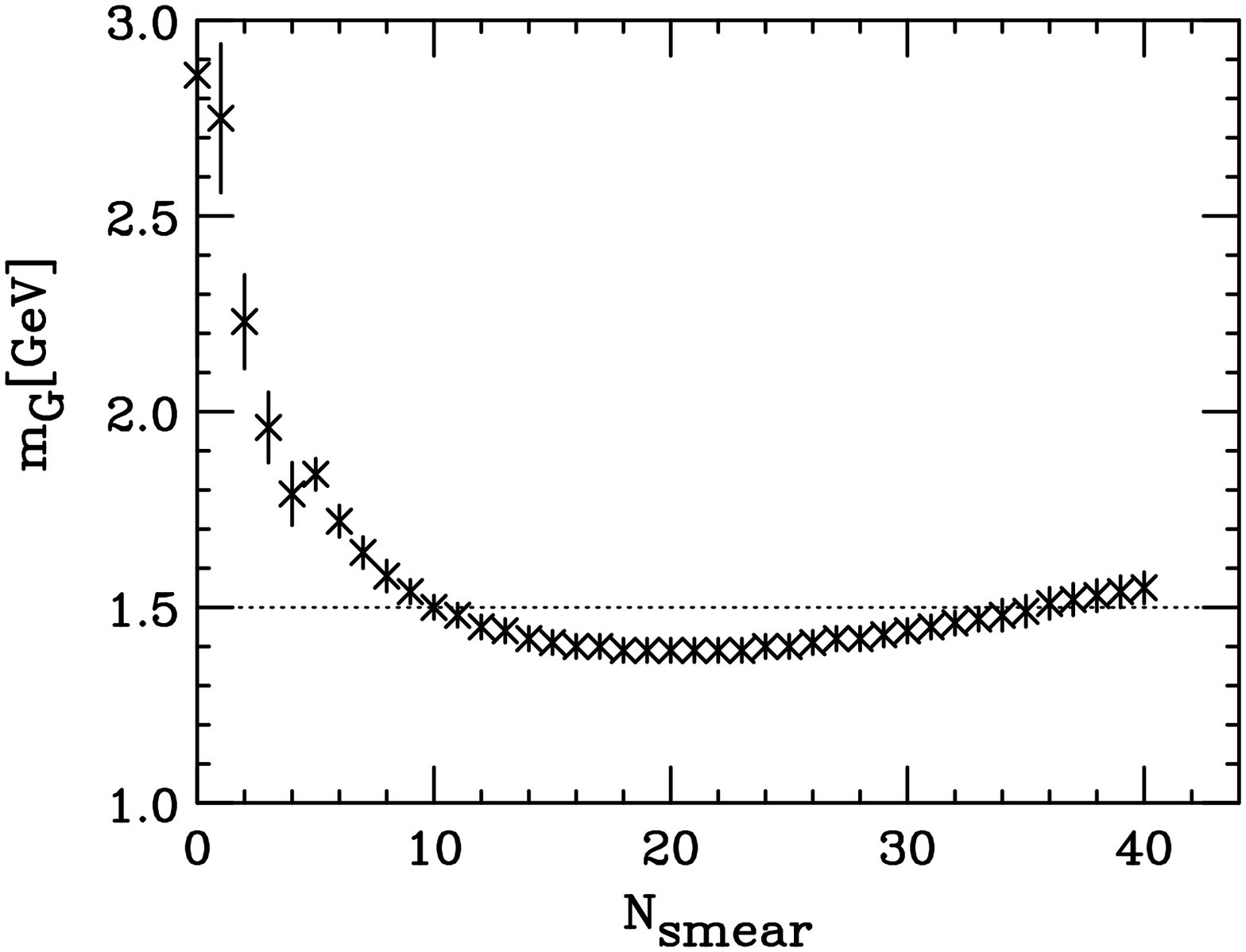,width=0.45\textwidth,angle=90}
&
\psfig{figure=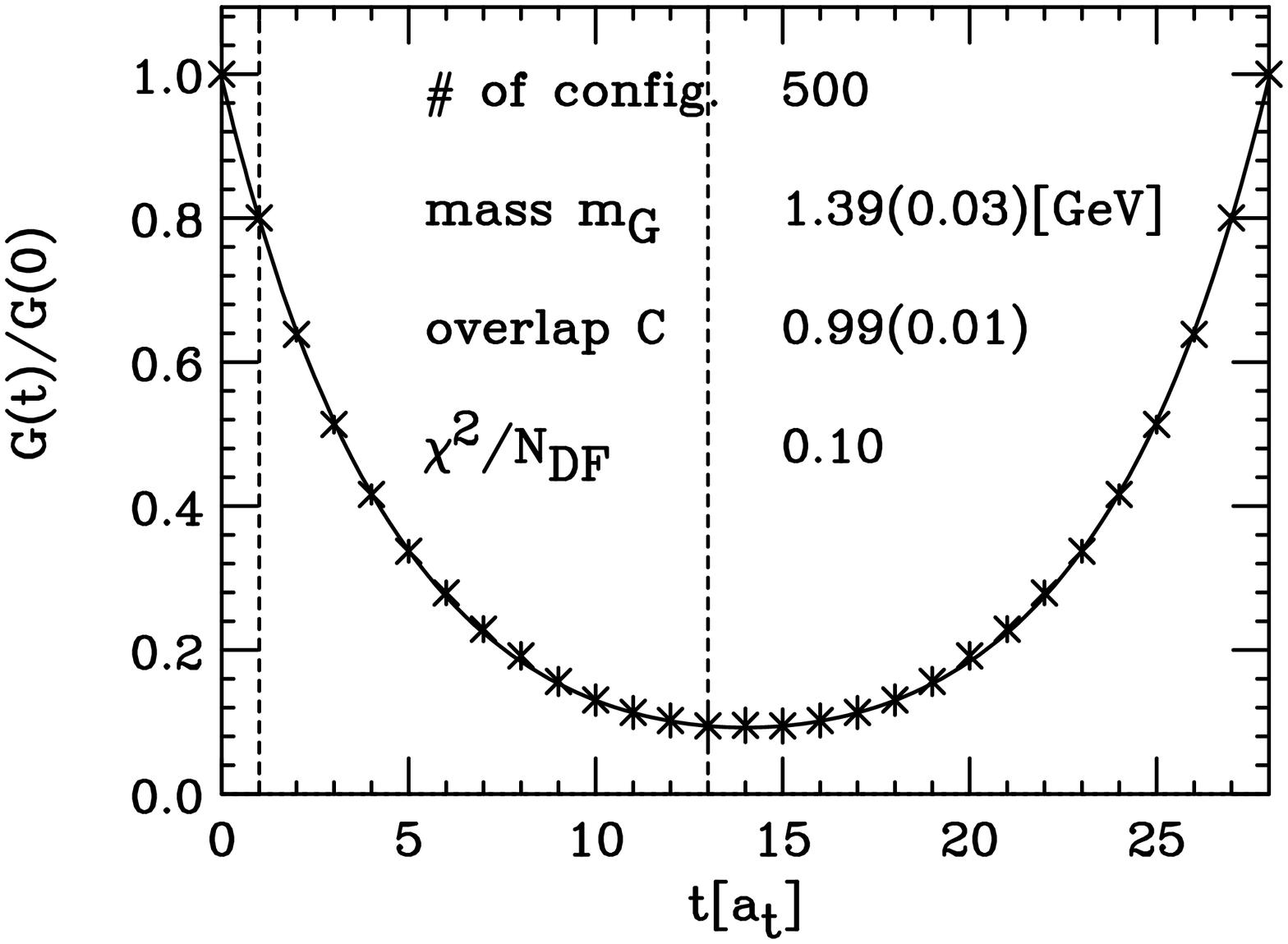,width=0.45\textwidth,angle=90}
\end{tabular}
\caption{The  fitted glueball  mass  at $T=0.87T_c$  is plotted  against
$\Nsmear$ (left figure).   The normalized glueball correator $G(t)/G(0)$
at $T=0.87T_c$  is plotted for  $\Nsmear=23$ (right figure).   The curve
denotes the  best hyperbolic cosine  fit with the interval  indicated by
the vertical dotted lines.}
\label{mass.correlator.t=0.87tc}
\end{figure}

At  finite  temperature, we  use  the  hyperbolic  cosine fitting  as  $
G(t)/G(0)  = C(e^{-m_{\rm G}t  a_t} +  e^{-m_{\rm G}(N_t  - t)  a_t}), $
where $N_t$ denotes  the number of the temporal  lattice size, and $a_t$
the  temporal lattice  spacing.   In \Fig{overlap.chi2.region.t=0.87tc},
the overlap $C$  and $\chi^2/\Ndf$ for $T=0.87 T_c$  are plotted against
$\Nsmear$.   In  the  whole  region,  we  have  reasonable  $\chi^2$  as
$\chi^2/\Ndf \alt 1$. The overlap  becomes maximum in the region $10 \le
\Nsmear    \le    23$.    In    this    region    as    is   shown    in
\Fig{mass.correlator.t=0.87tc},  the glueball  mass  becomes minimum  as
$m_{\rm G} = 1.39$ GeV in the interval $18 \le \Nsmear \le 23$.  We thus
obtain $m_{\rm G} = 1.39$ GeV at $T=0.87 T_c$.  The glueball size $R$ is
also estimated with \Eq{size} as $0.37$ fm $\,\alt\, R\, \alt\, 0.56$ fm
at $T=0.87T_c$  from the maximal overlap  region as $10  \le \Nsmear \le
23$.  We  show also  the glueball correlator  $G(t)$ at  $T=0.87T_c$ for
$\Nsmear=23$ in \Fig{mass.correlator.t=0.87tc}.
\section{Summary and discussions}
We have applied  the closed packing model to the  quenched QCD to derive
the relation  between the  mass and the  size of  the glueball so  as to
reproduce the  deconfinement phase  transition at $T_c  = 260$  MeV.  We
have then studied the mass and the size of the scalar glueball at finite
temperature by using $SU(3)$ lattice  QCD at the quenched level with the
anisotropic  lattice and  the smearing  method.  To  generate  the gauge
field  configurations, we  have  used the  tree-level Symanzik  $O(a^2)$
improved action with the lattice size $16^2\times 24\times N_t$, $N_t=96
(T=0)$ and  $N_t=28 (T=0.87T_c)$.  As  for the glueball size  $R(T)$, no
drastic change  has been observed  between $0.37$ fm  $\,\alt\, R(T=0)\,
\alt\, 0.57$ fm and $0.37$  fm $\,\alt\, R(T=0.87T_c)\, \alt\, 0.56$ fm.
We have obtained  the glueball mass $m_{\rm G} = 1.49$  GeV at $T=0$ and
$m_{\rm  G} = 1.39$  GeV at  $T=0.87T_c$.  This  result may  suggest the
slight reduction of the glueball mass at finite temperature, which would
provide a support  for our conjecture discussed in  Sect.~1 and Sect.~2.
However, this reduction does not seem  so drastic as is suggested by the
closed  packing model  for the  quenched QCD.   At any  rate, to  draw a
definite conclusion, much more statistics are necessary.
%
\\[1.2ex]
{\bf Acknowledgements}
\\[1.2ex]
The  calculations  have  been  performed   on  the  NEC  SX-5  in  Osaka
University.   The authors  thank  Dr. T.~Umeda  for  useful comments  and
discussions.

\end{document}